\magnification1200
\def\sk{Sherrington-Kirkpatrick}
\def\bra{\langle}
\def\ket{\rangle}
\def\qud{q_{12}}
\def\qdt{q_{23}}
\def\qut{q_{13}}
\def\qtq{q_{34}}

{\settabs 2\columns
\+Preprint n. &Dipartimento di Fisica\cr
\+&Universit\`a di Roma ``La Sapienza''\cr
\+&I.N.F.N. - Sezione di Roma\cr}
\vskip 2cm
\centerline{{\bf ABOUT THE OVERLAP DISTRIBUTION}
\footnote{}{Dedicated to the memory of Hiroomi Umezawa}\footnote{}{}}
\centerline{{\bf IN MEAN FIELD SPIN GLASS MODELS}\footnote{$\ddag$}{Research
supported in part by MURST (Italian Minister of University and Scientific and
Technological Research) and INFN (Italian National Institute for Nuclear
Physics).}}  \smallskip  \centerline{by} \smallskip \centerline
{Francesco Guerra}
\centerline{Dipartimento di Fisica, Universit\`a di Roma ``La Sapienza'',}
\centerline{and Istituto Nazionale di Fisica Nucleare, Sezione di Roma,}
\centerline{Piazzale Aldo Moro, 2, I-00185 Roma, Italy.}
\centerline{e-mail guerra@roma1.infn.it}
\centerline{WWW http://romagtc.roma1.infn.it/}
\bigskip\bigskip
\centerline{November 1995}
\vfill
\beginsection ABSTRACT.

We continue our presentation of mathematically rigorous results about the
Sherrington-Kirkpatrick mean field spin glass model. Here we establish some
properties of the distribution of overlaps between real replicas. They are in
full agreement with the Parisi accepted picture of spontaneous replica symmetry
breaking. As a byproduct, we show that the selfaveraging of the
Edwards-Anderson fluctuating order parameter, with respect to the external
quenched noise, implies that the overlap distribution is given by the
Sherrington-Kirkpatrick replica symmetric {\sl Ansatz}. This extends previous
results of Pastur and Shcherbina. Finally, we show how to generalize our
results to realistic short range spin glass models.
\vfill\eject

\beginsection 1. INTRODUCTION.

The physical content of the Sherrington-Kirkpatrick mean field spin glass model
[1,2] is fully contained in the probabilistic distribution of overlaps between
replicas of the system. In the approximate Sherrington-Kirkpatrick {\sl Ansatz}
[1], this distribution is concentrated on a numerical order parameter, while in
the more complete Parisi treatment [2], based on replica symmetry breaking, the
distribution shows ultrametric features and can be expressed through a
functional order parameter (see also [3,4,5]).

We have found a  general very elementary method able to generate rigorous
relations between replica averages, also in the infinite volume limit. The
method is based on simple analysis of fluctuations. The purpose of this paper
is to show some applications of this method.

If we consider four replicas, dubbed $1,2,3,4$, as defined in the following, 
then
the ultrametric picture gives a very simple statement about the joint
distributions of the overlaps $q_{12},q_{23}$ and $q_{12},q_{34}$,
respectively. Roughly speaking, for the joint
distribution of $q_{12},q_{23}$, it says that the distribution of $q_{23}$
is such that $q_{23}$ is equal to $q_{12}$ with probability one half, and
independent from $q_{12}$ with the same probability one half. Notice that
the statement does not depend on the precise form of the Parisi
functional order parameter, neither on the temperature. In this case the
overlaps $q_{12},q_{23}$ have one replica in common. If we consider
the overlaps $q_{12},q_{34}$, among disjoint replicas, then the
statement is the same as before, with the probabilities one half - one
half replaced by one third - two thirdies, respectively. The fact that
two overlaps, as $q_{12},q_{34}$, related to disjoint replicas, are
not completely independent is very peculiar to the ultrametric
picture, and is a consequence of the important phenomenon of the
absence of selfaveraging for some fluctuating observables.

As an example of application of our general method, we will give a
very simple rigorous proof for the following equalities among overlap
averages, direct consequences of the previous ultrametric statements,
$$\bra \qud^2 \qdt^2\ket={1\over 2}\bra \qud^4\ket+{1\over 2}{\bra
\qud^2\ket}^2, 
\eqno(1)$$
$$\bra \qud^2 \qtq^2\ket={1\over 3}\bra \qud^4\ket+{2\over 3}{\bra
\qud^2\ket}^2. 
\eqno(2)$$

In a recent paper [6], Marinari, Parisi, Ruiz-Lorenzo and Ritort    
have found,
among other things, convincing numerical evidence, based on computer simulations
on large systems, that these equalities hold also for spin glass systems
with short range interactions. This provides support to the idea that
some relevant features of the ultrametric picture survive also in the
transition from the mean field model to more realistic disordered short
range interactions (but see also [7]).

While it is true that the overlap distributions do not reproduce the
full physical content of the short range models, nevertheless it is
interesting to notice that our methods are able to establish equalities
(1,2), also in the short range case, modulo some technical assumptions
about the infinite volume limit, in full agreement with the computer
simulations.

As a byproduct of our analysis, we can also establish rigorously that
selfaveraging of the Edwards-Anderson fluctuating order parameter
implies that the overlap distribution is trivially given by the 
Sherrington-Kirkpatrick replica symmetric {\sl Ansatz}, through a
simple numerical order parameter. The precise statements are presented
in the following, but we can see immediately from (2) that
selfaveraging of the fluctuating order parameter would imply, in the
infinite volume limit, that
$$\bra \qud^2 \qtq^2\ket=\bra \qud^2\ket\bra \qtq^2\ket={\bra \qud^2\ket}^2,
\eqno(3)$$
and therefore 
$$\bra \qud^4\ket={\bra \qud^2\ket}^2.
\eqno(4)$$
In this case, all overlaps would be constant.

It is expected that for high temperatures and/or high values of the
external field the model exhibits selfaveraging (see for example [8]
and [3]), but for zero external field and low temperatures our result
shows that there can not be selfaveraging, because the
Sherrington-Kirkpatrick replica symmetric {\sl Ansatz} then becomes
incompatible with positivity of the entropy.

Our results, about lack of selfaveraging, complement previous similar
results by Pastur and Shcherbina [9] and Shcherbina [10], where
they consider fluctuating order parameters coming from the response
of the system to suitably chosen external random fields.

The paper is organized as follows. In Section 2, we recall the
basic definitions of the mean field spin glass model, and
introduce the overlap parameters and their averages, in connection
with the thermodynamic observables of the theory. In particular,
we show how to express the specific internal energy and its
temperature derivative, and the quadratic fluctuations for  the internal
energy and the order parameter in terms of overlap averages. In
Section 3, we exploit simple positivity properties in order to
prove some ultrametric inequalities. Then, we show that
selfaveraging for the fluctuating order parameter implies that
all overlaps are constant, according to the replica symmetric
{\sl Ansatz}. Section 4 is devoted to the proof of the metric
equalities. We will exploit the properties of selfaveraging for
the specific free energy and the internal energy. Finally,
Section 5 gives some conclusions and outlook for future
developments.

In conclusion, useful conversations with Roberto D'Autilia, Enzo
Marinari, Giorgio Parisi and Masha Shcherbina are gratefully
acknowledged. We would like to thank also Charles Newman and Dan 
Stein for making available Ref. [7] before publication, and useful
correspondence.
\bigskip
\beginsection 2. OVERLAP DISTRIBUTIONS.

The mean field spin glass model is defined on sites
$i=1,2,\dots,N$, to which Ising spin variables $\sigma_{i}=\pm
1$ are attached. The external quenched disorder is given by the
$N(N-1)/2$ independent and identical distributed random
variables $J_{ij}$, defined for each couple of sites. For the
sake of definiteness, we assume each $J_{ij}$ to be a centered
unit Gaussian with averages
$$E(J_{ij})=0,\quad E(J_{ij}^2)=1.
\eqno(5)$$
The Hamiltonian of the model is given by
$$H_N(\sigma,J)=-{1\over\sqrt{N}}\sum_{(i,j)}J_{ij}\sigma_i\sigma_j,
\eqno(6)$$
where the sum, extending to all spin couples, and the
normalizing factor ${1/\sqrt{N}}$ are typical of the mean field character of
the model.

For a given inverse temperature $\beta$, we introduce the
partition function $Z_{N}(\beta,J)$, the free energy per site
$f_{N}(\beta,J)$, the internal energy per site
$u_{N}(\beta,J)$, and the Boltzmann state $\omega_J$,
according to the definitions
$$Z_N(\beta,J)=\sum_{\sigma_1\dots\sigma_N}\exp(-\beta
H_N(\sigma,J)),
\eqno(7)$$
$$-\beta f_N(\beta,J)=N^{-1} \log Z_N(\beta,J),
\eqno(8)$$
$$\omega_{J}(A)=Z_N(\beta,J)^{-1}\sum_{\sigma_1\dots\sigma_N}A\exp(-\beta
H_N(\sigma,J)), 
\eqno(9)$$
$$u_N(\beta,J)=N^{-1}\omega_{J}(H_N(\sigma,J))=
\partial_{\beta}\bigl(\beta
f_N(\beta,J)\bigr), 
\eqno(10)$$
where $A$ is a generic function of the $\sigma$'s. In the
notation $\omega_J$, we have stressed the dependence of the Boltzmann
state on the external noise $J$, but, of course, there is also
a dependence on $\beta$ and $N$.

We are interested in the thermodynamic limit $N\to\infty$.

Let us consider a generic number $s$ of independent copies
(replicas) of the system, characterized by the Boltzmann
variables $\sigma^{(1)}_i$, $\sigma^{(2)}_i$, $\dots$,
distributed according to the product state
$$\Omega_J=\omega^{(1)}_J \omega^{(2)}_J \dots\omega^{(s)}_J,
\eqno(11)$$
where all $\omega^{(\alpha)}_J$ act on each one
$\sigma^{(\alpha)}_i$'s, and are subject to the {\sl
same} sample $J$ of the external noise.

The overlaps between two replicas $\alpha,\beta$ are
defined according to
$$Q_{\alpha\beta}(\sigma^{(\alpha)},\sigma^{(\beta)})={1\over
N}\sum_{i}\sigma^{(\alpha)}_i\sigma^{(\beta)}_i,
\eqno(12)$$
with the obvious bounds
$$-1\le Q_{\alpha\beta}\le 1.
\eqno(13)$$

For a generic smooth function $F$ of the overlaps, we
define the averages
$$E\Omega_J\bigl(F(Q_{12},Q_{13},\dots)\bigr),
\eqno(14)$$
where $\Omega_J$ acts on the $\sigma$ variables, and $E$
is the average with respect to the external noise $J$. We
introduce also random variables $q_{12}$, $q_{13}$, $\dots$,
through the definition of their averages                      
$$\bra F(q_{12},q_{13},\dots)\ket=E\Omega_J\bigl(F(Q_{12},Q_{13},\dots)\bigr).
\eqno(15)$$
We have made a careful distinction in notations. In fact, $\bra\ \ket$ contains
both Boltzmann averages $\Omega_J$ and noise averages $E$. 

It is important to
remark that the noise average $E$ introduces correlations between different
groups of replicas. For example, before the $E$ average, we have the
factorization  
$$\Omega_J (q_{12}^2
q_{34}^2)=(\omega_{J}^{(1)}\omega_{J}^{(2)})(q_{12}^2)\ 
(\omega_{J}^{(3)}\omega_{J}^{(4)})(q_{34}^2). 
\eqno(16)$$
But after the $E$ average we have in general
$$\bra \qud^2 \qtq^2\ket \ne \bra \qud^2\ket\bra \qtq^2\ket. 
\eqno(17)$$

The $\bra\ \ket$ 
averages are obviously invariant under permutations of the replicas, so that in
particular
$$\bra \qud^2\ket=\bra \qtq^2\ket.
\eqno(18)$$

Moreover, they are invariant under gauge transformations
$$q_{\alpha\beta}\to \epsilon_{\alpha}\epsilon_{\beta}q_{\alpha\beta}, 
\eqno(19)$$
where $\epsilon_{\alpha}=\pm 1$. This is an easy consequence of the fact that
each $\omega_{J}^{(\alpha)}$ is an even state on the respective $\sigma$'s.

There are important consequences from gauge invariance. First of all,
polynomials in the overlaps, which are not gauge invariant, have zero mean.
Moreover, the
distributions are uniquely defined by their restrictions to the case where all
values of the $q$'s are nonnegative. The general case, in all regions of
possible values for the $q$'s, is easily reconstructed using gauge invariance.

Overlap distributions carry the whole physical content of the theory. In fact,
the averages of the physical observables with respect to the external noise
can be expressed through averages involving overlap functions. As an example,
we give the following statements involving the internal energy, its $\beta$
derivative and its quadratic fluctuations.
\smallskip\noindent
{\bf Theorem 1.} The noise average of the internal energy is given through an
overlap average of the type
$$E\bigl(u_N(\beta,J)\bigr)=N^{-1}E\bigl(\omega_{J}(H_N(\sigma,J))\bigr)=
-{1\over 2}\beta(1-\bra\qud^2\ket).
\eqno(20)$$
\smallskip\noindent
{\bf Theorem 2.} The $\beta$ derivative of the noise average of the internal
energy is expressed as
$$\eqalignno{&\partial_{\beta}E\bigl(u_N(\beta,J)\bigr)
=-N^{-1}E\bigl(\omega_{J}(H_N^2)-\omega_{J}^2(H_N)\bigr)\cr
&=-{1\over 2}(1-\bra\qud^2\ket)+{{\beta}^2\over 2} N \bigl(\bra\qud^4\ket-
4 \bra\qud^2\qdt^2\ket+ 3 \bra\qud^2\qtq^2\ket\bigr). 
&(21)\cr}$$
\smallskip\noindent
{\bf Theorem 3.} For the quadratic fluctuation of the internal energy we have
$$\eqalignno{&E\bigl(u_N^2(\beta,J)\bigr)-\bigl(Eu_N(\beta,J)\bigr)^2=
N^{-2}\Bigl(E\bigl(\omega_{J}^2(H_N)\bigr)-
\bigl(E\omega_{J}(H_N)\bigr)^2\Bigr)\cr
&=-{1\over {2N}}\bra\qud^2\ket
-{1\over {2 N^2}}
+{3\over 2} {\beta}^2 (\bra\qud^2\qtq^2\ket-
 \bra\qud^2\qdt^2\ket)
+ {1\over 4} {\beta}^2 (\bra\qud^4\ket-{\bra\qud^2\ket}^2). 
&(22)\cr}$$

The proof of these results is very straigthforward, but it involves long
calculations. The main ingredient is nothing but integration by part with
respect to the external noise. In fact, for a unit Gaussian $J$ and a smooth
function $F$, we have
$$E\bigl(J F(J)\bigr)=E\bigl({\partial\over{\partial J}}F(J)\bigr).
\eqno(23)$$

As an example, we recall the proof of (20) (see [2], and [3]). Starting from
the definitions (6,9,10,12), we can write the following chain of equalities
$$\eqalignno{&-E\bigl(u_N(\beta,J)\bigr)=
-N^{-1}E\bigl(\omega_{J}(H_N(\sigma,J))\bigr)=
{1\over{N \sqrt{N}}} 
\sum_{(i,j)} E\bigl(J_{ij} \omega_J (\sigma_i \sigma_j)\bigr)\cr
&={\beta^2\over N^2} 
\sum_{(i,j)} \Bigl(1-E\bigl(\omega_J^2(\sigma_i \sigma_j)\bigr)\Bigr)
={\beta^2\over {2 N^2}}
\sum_{i,j} \Bigl(1-E\bigl(\omega_J^2(\sigma_i \sigma_j)\bigr)\Bigr)\cr
&={\beta^2\over 2}\Bigl(1-{1\over N^2}\sum_{i,j}
E\bigl(\omega_J^{(1)}(\sigma_i^{(1)} \sigma_j^{(1)}) 
\omega_J^{(2)}(\sigma_i^{(2)} \sigma_j^{(2)})\bigr)\Bigr)\cr
&={1\over 2}\beta(1-\bra\qud^2\ket).
&(24)\cr}$$

Notice that we have changed the sum over all couples $(i,j)$ into one half of
the sum over all $i$ and $j$, by taking into account that the diagonal
elements, corresponding to $i=j$, do not give any contribution. Formulae (21)
and (22) are established in a similar way. They involve a double integration by
parts, since the Hamiltonian $H_N$ appears quadratically.

We can also introduce the fluctuating order parameter
$$M_N^2(\beta,J)={1\over N^2} \sum_{i,j} \omega_J^2(\sigma_i \sigma_j)=
\Omega_J(Q_{12}^2),
\eqno(25)$$
and notice that
$$E\bigl(M_N^2(\beta,J)\bigr)=\bra\qud^2\ket,\quad 
E\bigl(M_N^4(\beta,J)\bigr)=\bra\qud^2\qtq^2\ket.
\eqno(26)$$
\bigskip
\beginsection{3. CONSEQUENCES OF SELFAVERAGING.}

Since (21) is clearly nonpositive, and (22) nonnegative, we have immediately
the following bounds
$$\bra\qud^2\qdt^2\ket\ge {1\over 4} \bra\qud^4\ket +
{3\over 4} \bra\qud^2\qtq^2\ket+O({1\over N}),
\eqno(27)$$  
$$\bra\qud^2\qdt^2\ket\le\bra\qud^2\qtq^2\ket+
{1\over 6} (\bra\qud^4\ket-{\bra\qud^2\ket}^2)+O({1\over N}).
\eqno(28)$$

Now we are ready to establish
\smallskip\noindent
{\bf Theorem 4.} The following estimates hold
$$\bra \qud^2 \qdt^2\ket\ge{1\over 2}\bra \qud^4\ket+{1\over 2}{\bra
\qud^2\ket}^2+O({1\over N}), 
\eqno(29)$$
$$\bra \qud^2 \qtq^2\ket\ge{1\over 3}\bra \qud^4\ket+{2\over 3}{\bra
\qud^2\ket}^2+O({1\over N}). 
\eqno(30)$$

In order to prove (30), it is sufficient to eliminate $\bra\qud^2\qdt^2\ket$
between (27) and (28). Then, (29) follows from the elimination of 
$\bra \qud^2 \qtq^2\ket$ between (30) and (27).

An obvious consequence of (29,30) is the following
\smallskip\noindent
{\bf Theorem 5.} In the limit for $N\to\infty$ we have
$$\liminf_{N\to\infty}\ \bigl(
\bra \qud^2 \qdt^2\ket-{1\over 2}\bra \qud^4\ket-{1\over 2}{\bra
\qud^2\ket}^2\bigr)\ge 0, 
\eqno(31)$$
$$\liminf_{N\to\infty}\ \bigl(
\bra \qud^2 \qtq^2\ket-{1\over 3}\bra \qud^4\ket-{2\over 3}{\bra
\qud^2\ket}^2)\bigr)\ge 0. 
\eqno(32)$$

Notice that the results of Theorem 4 and Theorem 5 rely only on positivity
properties for the expressions (21,22).

Finally, we can connect the quadratic fluctuations for the order parameter
defined in (25) with the quadratic fluctuations for the overlap $\qud^2$. In
fact, from (26) and (30), we immediately have
\smallskip\noindent
{\bf Theorem 6.} The following estimate holds
$$0\le\bra\qud^4\ket-{\bra\qud^2\ket}^2\le 
3 \Bigl(E\bigl(M_N^4(\beta,J)\bigr)-E^2\bigl(M_N^2(\beta,J)\bigr)\Bigr)
+O({1\over N}).
\eqno(33)$$

Therefore, we have a simple proof of the following very important result
\smallskip\noindent
{\bf Theorem 7.} Assume that the order parameter $M_N^2$ is selfaveraging in
the infinite volume limit, in the sense that for $N\to\infty$ along some
subsequence we have
$$\lim_{N\to\infty}
\Bigl(E\bigl(M_N^4(\beta,J)\bigr)-E^2\bigl(M_N^2(\beta,J)\bigr)\Bigr)=0.
\eqno(34)$$
Then, along the same subsequence
$$\lim_{N\to\infty}    
(\bra\qud^4\ket-{\bra\qud^2\ket}^2)=0.
\eqno(35)$$
If, moreover, we chose $N\to\infty$ along a subsequence such that
$$\lim_{N\to\infty}\bra\qud^2\ket={\overline q}^2,
\eqno(36)$$
where for example
$${\overline q}^2=\limsup_{N\to\infty}\ \bra\qud^2\ket,
\eqno(37)$$
then along the same subsequence we have for any gauge invariant function $F$
$$\lim_{N\to\infty} \bra F(\qud, \qut, \dots)\ket=
F({\overline q},{\overline q},\dots),
\eqno(38)$$
so that the overlaps have the \sk\ replica symmetric form.

Clearly, (35) follows from (38) through (33), while (38) is a simple
consequence of the absence of dispersion for the overlaps.

This result complements previous results by Pastur and Shcherbina [9], and
Shcherbina [10], where they prove that quadratic selfaveraging for some order
parameters, coming from the response of the system to random external fields,
implies the \sk\ form for the solution of the model.
\bigskip
\beginsection{4. THE ULTRAMETRIC EQUALITIES.}

In [9], Pastur and Shcherbina have proven the selfaveraging of the free energy
per site in quadratic mean, in the infinite volume limit. In fact, by a mild
improvement of their methods, one can easily show [11] the following
\smallskip\noindent
{\bf Theorem 8.} The quadratic fluctuation of the free energy per site can be
estimated as follows
$$E\bigl((N^{-1} \log Z_N(\beta,J))^2\bigr)
-\bigl(E(N^{-1} \log Z_N(\beta,J))\bigr)^2\le
{\beta^2\over {2 N}}\bra \qud^2\ket+O({1\over N^2}).
\eqno(39)$$

By using standard techniques of statistical mechanics, one can easily derive
statements about the quadratic fluctuation of the internal energy. A precise
formulation is the following.

Let $N$ go to infinity along a subsequence such that the average specific energy
is convergent
$$\lim_{N\to\infty} N^{-1} E \log Z_N(\beta,J))= -\beta f(\beta).
\eqno(40)$$

Due to convexity in $\beta$ the subsequence can be chosen without exploiting
compacteness or the axiom of choice [11], as for any subsequence in this paper,
by working for example always with $\limsup$'s for a denumerable dense set of
$\beta$'s.

Since the function $-\beta f(\beta)$ is convex in $\beta$, there exist the
left and right $\beta$ derivatives $u_{(\pm)}(\beta)$, and, moreover, they are
equal for all values of $\beta$, with the possible exclusion of a set of zero
Lebesgue measure. Then, we have [11]
\smallskip\noindent
{\bf Theorem 9.} For $N$ going to infinity, along the chosen subsequence, the
quadratic fluctuation of the specific internal energy satisfies
$$\limsup_{N\to\infty}
\Bigl(E\bigl(u_N^2(\beta,J)\bigr)-\bigl(Eu_N(\beta,J)\bigr)^2\Bigr)\le
{1\over 4}\bigl(u_{(+)}(\beta)-u_{(-)}(\beta)\bigr)^2.
\eqno(41)$$
Therefore, we have quadratic selfaveraging for almost all values of $\beta$.

For a complete detailed proof of Theorem 8 and Theorem 9, we refer to our
forthcoming review paper [11].

>From now on, we let $N\to\infty$ along subsequences such that not only the
specific free energy but also the averages $\bra\qud^2\ket$, $\bra\qud^4\ket$,
$\bra\qud^2 \qdt^2\ket$, $\bra\qud^2 \qtq^2\ket$ have well definite limits.

An immediate consequence of Theorem 3 and Theorem 9 is the following
\smallskip\noindent
{\bf Theorem 10.} In the limit $N\to\infty$ along the stated subsequence, for
almost all values of $\beta$, we have
$$\bra\qud^2 \qtq^2\ket- \bra\qud^2 \qdt^2\ket=
{1\over 6} ( \bra\qud^4\ket - {\bra\qud^2\ket}^2).
\eqno(42)$$

Further information can be gathered from (20,21). In fact, at fixed $N$, by
taking the $\beta$ derivative in (20), and by exploiting (21), we have
$$-{1\over 2}+{1\over 2}\beta \partial_{\beta}\bra\qud^2\ket=
{\beta^2 \over 2} N A_N^2(\beta),
\eqno(43)$$
where we have defined the nonnegative $A_N^2(\beta)$ through
$$A_N^2(\beta)=\bra\qud^4\ket-
4 \bra\qud^2\qdt^2\ket+ 3 \bra\qud^2\qtq^2\ket
-{1\over {N\beta^2}}(1-\bra\qud^2\ket).
\eqno(44)$$

Let us integrate (43) in $d\beta^2$ on the generic interval $[\beta^2_1,
\beta^2_2 ]$, with
$0<\beta^2_1<\beta^2_2$. We have
$$\int^{\beta^2_2}_{\beta^2_1} A_N^2(\beta) d\beta^2=
{2\over N} 
\bigl({\bra\qud^2\ket}_{\beta^2_2}-{\bra\qud^2\ket}_{\beta^2_1}\bigr)
-{1\over N} \log {\beta^2_2 \over \beta^2_1}.
\eqno(45)$$

Therefore, we have that $A_N^2(\beta)$ converges to zero in quadratic
mean in any $\beta^2$ interval. We can extract a subsequence such that 
$A_N^2(\beta)$ converges to zero, for all values of $\beta$, by excluding
a set of zero Lebesgue measure. By recalling the definition (44), we then
arrive at
\smallskip\noindent
{\bf Theorem 11.} In the limit $N\to\infty$ along the stated subsequence,
for almost all values of $\beta$, we have
$$\bra\qud^4\ket-
4 \bra\qud^2\qdt^2\ket+ 3 \bra\qud^2\qtq^2\ket=0.
\eqno(46)$$

By collecting all results of Theorems 10 and 11, we can establish the following
equalities, in full agreement with the Parisi ultrametric solution of the mean
field spin glass model.
\smallskip\noindent
{\bf Theorem 12.} For almost all values of $\beta$, and for $N$ going to
infinity along a chosen subsequence, we have in the limit
$$\bra \qud^2 \qdt^2\ket={1\over 2}\bra \qud^4\ket+{1\over 2}{\bra
\qud^2\ket}^2, 
\eqno(47)$$
$$\bra \qud^2 \qtq^2\ket={1\over 3}\bra \qud^4\ket+{2\over 3}{\bra
\qud^2\ket}^2. 
\eqno(48)$$

Therefore, we have proven the main result of this paper. The possible
restrictions on the values of $\beta$ and the limiting subsequences are a
consequence of the method of proof. In fact, we believe that all results are
true for any value of $\beta$ and $N$ going to infinity without restrictions.
But we would need a better control on the limits.

It is immediate to generalize the results of Theorem 12 to models with short
range interaction. In fact, we can add to the interaction Hamiltonian of the
short range  model a small mean field interaction of the type $\lambda H_N$,
with $H_N$ defined as in (6). By taking $\lambda\to 0$, after the infinite
volume limit, we still have (47,48). The general validity of (47,48) is also
confirmed by numerical simulations [6].
\bigskip
\beginsection{5. CONCLUSIONS AND OUTLOOK FOR FUTURE DEVELOPMENTS.}

In conclusion, we can see that all results explained in this paper rely on very
elementary arguments, based essentially on positivity and convexity. These
results are easily generalized.

For example, we can  add a small interaction of the form
$$\lambda\sum_i J_i \sigma_i,
\eqno(49)$$
where the $J_i$ are independent centered unit Gaussian external noises. By
repeating all arguments in this paper, we immediately establish equalities of
the type
$$\bra \qud \qdt \ket={1\over 2}\bra \qud^2\ket+{1\over 2}{\bra
\qud\ket}^2, 
\eqno(47)$$
$$\bra \qud \qtq\ket={1\over 3}\bra \qud^2\ket+{2\over 3}{\bra
\qud\ket}^2. 
\eqno(48)$$

In a forthcoming paper, we will show other applications of this method.
\vfill\eject\bigskip
\beginsection REFERENCES

\item{ [1]} D. Sherrington and S. Kirkpatrick: Solvable model of a spin glass,
Phys. Rev. Lett., {\bf35}, 1792 (1975).
\item{ [2]} M. M\'ezard, G. Parisi, and M. A. Virasoro: {\sl Spin Glass Theory
and Beyond}, World Scientific, Singapore, 1987, and reprints included there.
\item{ [3]} F. Guerra: Fluctuations and Thermodynamic Variables in Mean
Field Spin Glass Models, in: {\sl Stochastic Processes, Physics and
Geometry}, S. Albeverio {\sl et al.}, eds, World Scientific, Singapore, 1995.
\item{ [4]} F. Guerra: Functional Order Parameters for the Quenched Free
Energy in Mean Field Spin Glass Models, Preprint, Rome, 1992, in press.
\item{ [5]} F. Guerra: The Cavity Method in the Mean Field Spin Glass Model.
Functional Representations of Thermodynamic Variables,
in: {\sl Advances in Dynamical Systems and Quantum Physics}, S. Albeverio {\sl 
et al.},
eds, World Scientific, Singapore, 1995.
\item{ [6]} E. Marinari, G. Parisi, J. Ruiz-Lorenzo and F. Ritort, Numerical
Evidence for Spontaneously Broken Replica Symmetry in $3D$ Spin Glasses, to
appear.
\item{ [7]} C. Newman and D. Stein, Non-Mean-Field Behavior of Realistic
Spin Glasses, to appear. \item{ [8]} M. V. Shcherbina, in preparation.
\item{ [9]} L. A. Pastur and M. V. Shcherbina: The Absence of
Self-Averaging of the Order Parameter in the \sk\ Model, J. Stat. Phys., {\bf
62}, 1 (1991).
\item{ [10]} M. V. Scherbina: More about Absence of
Selfaverageness of the Order Parameter in the \sk\ Model, CARR Reports in 
Mathematical Physics, n.
3/91, Department of Mathematics, University of Rome ``La Sapienza'', 1991.
\item{[11]} F. Guerra: On the mean field spin glass model, in preparation.
\vfill\eject\bye